\documentclass[a4paper,12pt,titlepage]{article}

\usepackage{amsthm}
\usepackage{amsmath,amssymb,revsymb}
\usepackage[dvipdfmx]{graphicx}
\usepackage{here}
\usepackage{bm}
\usepackage{fancybox}
\usepackage{secdot}
\usepackage{comment}
\usepackage{braket}
\usepackage{titlesec}
\usepackage{geometry}
\usepackage{hyperref}
\usepackage{slashed}
\usepackage{fancyhdr}
\usepackage{boxedminipage}
\usepackage[dvipdfmx]{color}
\usepackage{mathrsfs}
\usepackage{feynmf}
\usepackage{cite}

\titleformat*{\section}{\LARGE\bfseries}
\titleformat*{\subsection}{\Large\bfseries}

\hypersetup{
setpagesize=false,
 bookmarksnumbered=true,%
 bookmarksopen=true,%
 colorlinks=true,%
 linkcolor=blue,
 citecolor=blue,
 }
 
\makeatletter 
\long\def\@makefntext#1{\parindent 1em\noindent 
\@hangfrom{\hbox to 1.8em{\hss$^{\@thefnmark}$}}#1}
\makeatother

\begin{document}

\fancypagestyle{foot}
{
\fancyfoot[L]{$^{*}$E-mail address : yakkuru$\_$111@ruri.waseda.jp\\
$^{\dagger}$E-mail address : msato@hirosaki-u.ac.jp\\
$^{\ddagger}$E-mail address : h20ms117@hirosaki-u.ac.jp}
\fancyfoot[C]{}
\fancyfoot[R]{}
\renewcommand{\headrulewidth}{0pt}
\renewcommand{\footrulewidth}{0.5pt}
}

\renewcommand{\footnoterule}{%
  \kern -3pt
  \hrule width \columnwidth
  \kern 2.6pt}

%%%%%%%%%%%%%%%%%%% TITLE PAGE %%%%%%%%%%%%%%%%%%%%%%%%%%%%%

\begin{titlepage}
\begin{flushright}
\begin{minipage}{0.2\linewidth}
\normalsize
\end{minipage}
\end{flushright}

\begin{center}

\vspace*{5truemm}
\Large
\bigskip\bigskip

\LARGE\textbf{Superstring Backgrounds in String Geometry}%

\Large

\bigskip\bigskip
Masaki Honda$^{1,*}$, Matsuo Sato$^{2,\dagger}$ and Taiki Tohshima$^{2, \ddagger}$
\vspace{1cm}

{\large$^{1}$ \it{Department of Physics, Waseda University, 
Totsuka-cho 1-104, Shinjuku-ku, Tokyo 169-8555, Japan}}\\
{\large$^{2}$ \it{Department of Mathematics and Physics, Graduate School of Science and Technology, Hirosaki University, Bunkyo-cho 3, Hirosaki, Aomori 036-8561, Japan}}\\
\bigskip\bigskip\bigskip
\large\textbf{Abstract}\\
\end{center}
String geometry theory is a candidate of the non-perturbative formulation of string theory. In order to determine the string vacuum, we need to clarify how superstring backgrounds are described in string geometry theory. In this paper, we show that all the type IIA, IIB, SO(32) type I, and   SO(32)  and $E_8 \times E_8$ heterotic superstring backgrounds are embedded in configurations of the fields of a single string geometry model. Especially, we show that the configurations satisfy the equations of motion of the string geometry model in $\alpha' \to 0$ if and only if the embedded string backgrounds satisfy the equations of motion of the supergravities, respectively. This means that classical dynamics of the string backgrounds are described as a part of classical dynamics in string geometry theory. Furthermore, we define an energy of the configurations in the string geometry model because they do not depend on the string geometry time. A string background can be determined by minimizing the energy.

\thispagestyle{foot} 

\end{titlepage}

\baselineskip 7.5mm

%%---------------------- CONTENTS ------------------------------ %%

\tableofcontents

%%---------------------- MAIN ------------------------------ %%

\parindent=20pt

\jot=8pt

\renewcommand\thefootnote{\textcolor{red}{\arabic{footnote}}}

\setlength{\abovedisplayskip}{12pt} 
\setlength{\belowdisplayskip}{12pt} 

%--------section1-------------%
\section{Introduction}
Superstring theory is a promising candidate of a unified theory including gravity. However, superstring theory is established at only the perturbative level as of this moment. The perturbative superstring theory lacks predictability because it has many perturbatively stable vacua.

String geometry theory is a candidate of non-perturbative formulation of superstring theory~\cite{Sato:2017qhj}, which can determine a non-perturbatively stable vacuum. In string geometry theory, the path-integral of the perturbative superstring theory on the flat string background is derived by taking a Newtonian limit of fluctuations around a fixed flat background in an Einstein-Hilbert action coupled with any field on string manifolds~\cite{Sato:2017qhj, Sato:2020szq}\footnote{A perturbative topological string theory is also derived from the topological sector of string geometry theory \cite{Sato:2019cno}. }. That is, the spectrum and all order scattering amplitudes in superstring theory on a flat background are derived from string geometry theory. 
However, perturbative string theory describes only propagation and interactions of strings in a fixed classical string background, and cannot describe dynamics of the classical string background itself. Only the consistency with the Weyl invariance requires that the string background satisfies the equations of motion of supergravity. That is, string backgrounds are treated as external fields in the perturbative string theory. In order to determine a string background, a non-perturbative string theory needs to be able to describe dynamics of the string backgrounds not in consequence of consistency.

In paper \cite{Honda:2020sbl}, as a first step to determine the string vacuum, the authors studied how arbitrary configurations of the bosonic string backgrounds are embedded in configurations of the fields of a bosonic string geometry model. Especially, the authors showed that the action of the string backgrounds is obtained by a consistent truncation of the action of the string geometry model; the configurations of the fields of string geometry model satisfy their equations of motion if and only if the embedded configurations of the string backgrounds satisfy their equations of motion. This means that classical dynamics of the string backgrounds are described as a part of classical dynamics in string geometry theory. This fact supports the conjecture that string geometry theory is a non-perturbative formulation of string theory. 

This truncation is valid without taking $\alpha' \to 0$ limit, which corresponds to $X (\bar{\sigma}) \to x$. This fact will be important to derive the path-integral of the non-linear sigma model from fluctuations around the string background configurations in the string geometry theory, since the string backgrounds in  the non-linear sigma model depend not only on the string zero modes $x$ but also on the other modes of $X(\sigma)$ \cite{Callan:1985ia, Fradkin:1985ys, Polchinski:1998rr}:
\begin{align}
\label{heterotic worldsheet action}
S=\frac{1}{4 \pi \alpha'} \int d^2 \sigma \sqrt{g} \left[ \left( g^{ab} G_{\mu \nu}(X(\sigma)) + i \epsilon^{ab} B_{\mu \nu}(X(\sigma))  \right) \partial_{a}X^{\mu} \partial_{b} X^{\nu} + \alpha' R \phi(X(\sigma))  \right].
\end{align}

In this paper, we generalize the results in \cite{Honda:2020sbl} to the supersymmetric case, which possesses interesting problems as follows. In general, it is too difficult to define non-linear sigma models in R-R backgrounds in the NS-R formalism of string theory.  On the other hand, if one can perform a supersymmetric generalization of the above results as they are, there is an apparent contradiction that we can derive non-linear sigma models in R-R backgrounds in the NS-R formalism from string geometry theory. We will see how this contradiction is resolved.  There is an another interesting problem: Chern-Simons terms cannot be defined in string geometry models because they are infinite dimensional, although supergravities, which should be reproduced from the models, possess Chern-Simons terms. Nevertheless, we will see that the type  IIA and IIB supergravities are reproduced from a string geometry model. 

The organization of this paper is as follows. In Sec.~2, we introduce a string geometry model. In Sec.~3, we identify type IIA and IIB string background configurations and obtain the equations of motions of the type IIA and IIB supergravities from the equations of motions of the string geometry model by a consistent truncation in $\alpha' \to 0$ limit. In Sec.~4, we define an energy of the string background configurations because they do not depend on the string geometry time. A string background can be determined by minimizing the energy. In Sec.~5,  all the five supergravities in the ten-dimensions are derived from a single string geometry model by consistent truncations.  In Sec.~6, we conclude and discuss our results.

%--------section2-------------%
\section{String geometry model}
We study a string geometry model\footnote{The action of string geometry theory is not determined as of this moment. On this stage, we should consider various possible actions. Then, we call each action a {\it string geometry model} and call the whole formulation {\it string geometry theory}. In \cite{Sato:2017qhj}, the perturbative superstring theory on the flat spacetime is derived from a gravitational model coupled with a $u(1)$ field on a Riemannian string manifold, whereas in \cite{Sato:2020szq}, it is derived from gravitational models coupled with arbitrary fields on a  Riemannian string manifold. Thus, the perturbative superstring theory on the flat spacetime is derived from this model.} whose action is given by,
\begin{eqnarray}
S=\frac{1}{G_N}\int \mathcal{D}\bold{E} \mathcal{D}\bar{\tau} \mathcal{D}\bold{X}_{\hat{D}} 
\sqrt{\bold{G}} \left( e^{-2 \Phi} \left( \mathbf{R}  + 4 \nabla_{\bold{I}} \Phi \nabla^{\bold{I}} \Phi - \frac{1}{2} |\mathbf{H} |^{2}   \right) -\frac{1}{2}\sum_{p=1}^9
|\tilde{\bold{F}}_p|^2     
\right),
\label{action of bos string-geometric model}
\end{eqnarray}
where $G_{N}$ is a constant, $\bold{I}=\{d,(\mu \bar{\sigma} \bar{\theta}) \}$, $| \mathbf{H} |^{2}:= \frac{1}{3!} \mathbf{G}^{\bold{I}_{1} \bold{J}_{1}} \mathbf{G}^{\bold{I}_{2} \bold{J}_{2}} \mathbf{G}^{\bold{I}_{3} \bold{J}_{3}} \mathbf{H}_{\bold{I}_{1} \bold{I}_{2} \bold{I}_{3}} \mathbf{H}_{\bold{J}_{1} \bold{J}_{2} \bold{J}_{3}}$, and we use the Einstein notation for the index $\bold{I}$. The action~(\ref{action of bos string-geometric model}) consists of  a scalar curvature    $\mathbf{R}$ of a metric $\mathbf{G}_{\bold{I}_{1} \bold{I}_{2}}$, a scalar field $\Phi$, a field strength $\mathbf{H}_{ \bold{I}_{1} \bold{I}_{2} \bold{I}_{3} }$ of a two-form field $\mathbf{B}_{\bold{I}_{1} \bold{I}_{2}}$ and $\tilde{\bold{F}}_p$. $\tilde{\bold{F}}_p$ are defined by 
$\sum_{p=1}^9 \tilde{\bold{F}}_p=e^{-\bold{B}_2}\wedge \sum_{k=1}^9 \bold{F}_k$, where $\bold{F}_k$ are field strengths of (k-1)-form fields $\bold{A}_{k-1}$. For example, $\tilde{\bold{F}}_5
=
\bold{F}_5
-\bold{B}_2\wedge 
\bold{F}_3
+\frac{1}{2}\bold{B}_2\wedge \bold{B}_2\wedge \bold{F}_1$. 
They are defined on a {\it Riemannian string manifold}, whose definition is given in \cite{Sato:2017qhj}. 
String manifold is constructed by patching open sets in string model space $E$, whose definition is summarized as follows. 
First, a global time $\bar{\tau}$ is defined canonically and uniquely on a super Riemann surface $\bar{\bold{\Sigma}}$ by the real part of the integral of an Abelian differential uniquely defined on $\bar{\bold{\Sigma}}$ \cite{Krichever:1987a, Krichever:1987b}.
We restrict $\bar{\bold{\Sigma}}$ to a $\bar{\tau}$ constant line and obtain $\bar{\bold{\Sigma}}|_{\bar{\tau}}$. An embedding of $\bar{\bold{\Sigma}}|_{\bar{\tau}}$ to $\mathbb{R}^{d}$ represents a many-body state of superstrings in $\mathbb{R}^{d}$, and is parametrized by coordinates $(\bar{\bold{E}}, \bold{X}_{\hat{D}_{T}}(\bar{\tau}), \bar{\tau})$\footnote{$\,\, \bar{}$ represents a representative of the super diffeomorphism and super Weyl transformation on the worldsheet. Giving a super Riemann surface $\bar{\bold{\Sigma}}$ is equivalent to giving a  supervierbein $\bar{\bold{E}}$ up to super diffeomorphism and super Weyl transformations.} where $\bar{\bold{E}}$ is a super vierbein on  $\bar{\bold{\Sigma}}$ and $\bold{X}_{\hat{D}_{T}}(\bar{\tau})$ is a map from  $\bar{\bold{\Sigma}}|_{\bar{\tau}}$ to $\mathbb{R}^{d}$.  $\hat{D}_T$ represents all the backgrounds except for the target metric, that consist of the B-field, the dilaton and the R-R fields. String model space $E$  is defined by the collection of the string states by considering all the  $\bar{\bold{\Sigma}}$, all the values of $\bar{\tau}$, and all the $\bold{X}_{\hat{D}_{T}}(\bar{\tau})$. How near the two string states is defined by how near the values of $\bar{\tau}$ and how near $\bold{X}_{\hat{D}_{T}}(\bar{\tau})$ \footnote{$\bar{\bold{E}}$  is a discrete variable in the topology of string geometry, where an $\epsilon$-open neighborhood of $[\bar{{\boldsymbol \Sigma}}, \bold{X}_{\hat{D}_{T}s}(\bar{\tau}_s), \bar{\tau}_s]$ is defined by
\begin{eqnarray}
U([\bar{\bold{E}}, \bold{X}_{\hat{D}_{T}s}(\bar{\tau}_s), \bar{\tau}_s], \epsilon)
:=
\left\{[\bar{\bold{E}}, \bold{X}_{\hat{D}_{T}}(\bar{\tau}), \bar{\tau}] \bigm|
\sqrt{|\bar{\tau}-\bar{\tau}_s|^2
+\| \bold{X}_{\hat{D}_{T}}(\bar{\tau}) -\bold{X}_{\hat{D}_{T} s}(\bar{\tau}_s) \|^2}
<\epsilon   \right\}, \label{SuperNeighbour}
\end{eqnarray}
As a result, $d \bar{\bold{E}}$  cannot be a part of basis that span the cotangent space in (\ref{cotangen}), whereas fields are functionals of $\bar{\bold{E}}$ as in (\ref{LineElement}). The precise definition of the string topology is given in the section 2 in \cite{Sato:2017qhj}.}.  By this definition, arbitrary two string states on a connected super Riemann surface in $E$ are connected continuously. Thus, there is an one-to-one correspondence between a super Riemann surface in $\mathbb{R}^{d}$ and a curve  parametrized by $\bar{\tau}$ from $\bar{\tau}=-\infty$ to $\bar{\tau}=\infty$ on $E$. That is, curves that represent asymptotic processes on $E$ reproduce the right moduli space of the super Riemann surfaces in $\mathbb{R}^{d}$. Therefore, a string geometry model possesses all-order information of superstring theory.  Actually, all order perturbative scattering amplitudes  of the   superstrings in the flat spacetime are derived from the string geometry theory as in \cite{Sato:2017qhj, Sato:2020szq}\footnote{The consistency of the perturbation theory determines $d=10$ (the critical dimension).}. 
The cotangent space is spanned by 
\begin{eqnarray}
d \bold{X}^{d}_{\hat{D}_{T}} &:=& d \bar{\tau}
\nonumber \\
d \bold{X}^{(\mu \bar{\sigma} \bar{\theta}) }_{\hat{D}_{T}}&:=& d \bold{X}^{\mu}_{\hat{D}_{T}} \left( \bar{\sigma}, \bar{\tau}, \bar{\theta} \right), \label{cotangen}
\end{eqnarray}
where $\mu=1, \dots, 10$. 
The summation over $(\bar{\sigma}, \bar{\theta})$ is defined by 
$\int d\bar{\sigma}d^2\bar{\theta} \hat{\bold{E}}(\bar{\sigma}, \bar{\tau}, \bar{\theta})$.
$\hat{\bold{E}}(\bar{\sigma}, \bar{\tau}, \bar{\theta})
:=
\frac{1}{\bar{n}}\bar{\bold{E}}(\bar{\sigma}, \bar{\tau}, \bar{\theta})$, where $\bar{n}$ is the lapse function of the two-dimensional metric.
This summation is transformed as a scalar under $\bar{\tau} \mapsto \bar{\tau}'(\bar{\tau}, \bold{X}_{\hat{D}_T}(\bar{\tau}))$ and invariant under a supersymmetry transformation $(\bar{\sigma}, \bar{\theta}) \mapsto (\bar{\sigma}'(\bar{\sigma}, \bar{\theta}), \bar{\theta}(\bar{\sigma}, \bar{\theta}))$. 
As a result, the action (\ref{action of bos string-geometric model}) is invariant under this $\mathcal{N}=(1,1)$ supersymmetry transformation.
%a supersymmetry transformation,
%\begin{equation}
%(\bar{\sigma}, \bar{\theta}) \mapsto (\bar{\sigma}'(\bar{\sigma}, \bar{\theta}), \bar{\theta}(\bar{\sigma}, \bar{\theta})). \label{supertrans}
%\end{equation}
An explicit form of the line element is given by
\begin{eqnarray}
&&ds^2(\bar{\bold{E}}, \bold{X}_{\hat{D}_{T}}(\bar{\tau}), \bar{\tau}) \nonumber \\
=&&G(\bar{\bold{E}}, \bold{X}_{\hat{D}_{T}}(\bar{\tau}), \bar{\tau})_{dd} (d\bar{\tau})^2 \nonumber \\ 
&&+2 d\bar{\tau} \int d\bar{\sigma} d^2\bar{\theta}  \hat{\bold{E}} \sum_{\mu} G(\bar{\bold{E}}, \bold{X}_{\hat{D}_{T}}(\bar{\tau}), \bar{\tau})_{d \; (\mu \bar{\sigma} \bar{\theta})} d \bold{X}_{\hat{D}_{T}}^{\mu}(\bar{\sigma}, \bar{\tau}, \bar{\theta}) \nonumber \\
&&+\int d\bar{\sigma} d^2\bar{\theta} \hat{\bold{E}}  \int d\bar{\sigma}'  d^2\bar{\theta}' \hat{\bold{E}}'  \sum_{\mu, \mu'} G(\bar{\bold{E}}, \bold{X}_{\hat{D}_{T}}(\bar{\tau}), \bar{\tau})_{ \; (\mu \bar{\sigma} \bar{\theta})  \; (\mu' \bar{\sigma}' \bar{\theta}')} d \bold{X}_{\hat{D}_{T}}^{\mu}(\bar{\sigma}, \bar{\tau}, \bar{\theta}) d \bold{X}_{\hat{D}_{T}}^{\mu'}(\bar{\sigma}', \bar{\tau}, \bar{\theta}'). 
\nonumber \\
&& \label{LineElement}
\end{eqnarray}
The inverse metric $\mathbf{G}^{\bold{I}\bold{J}}(\bar{\bold{E}}, \bold{X}_{\hat{D}_{T}}(\bar{\tau}), \bar{\tau})$\footnote{Like this, the fields $\mathbf{G}_{\bold{I}\bold{J}}$, $\Phi$, $\mathbf{B}_{\bold{L}_1\bold{L}_2}$ and $\bold{A}_{\bold{L}_1 \cdots \bold{L}_{p-1}}$ are functionals of the coordinates $\bar{\bold{E}}$, $\bold{X}_{\hat{D}_{T}}(\bar{\tau})$ and $\bar{\tau}$.} is defined by $\mathbf{G}_{\bold{I}\bold{J}}\mathbf{G}^{\bold{J}\bold{K}}=\mathbf{G}^{\bold{K}\bold{J}}\mathbf{G}^{\bold{J}\bold{I}}=\delta^{\bold{K}}_{\bold{I}}$, where $\delta^{d}_{d}=1$ and $\delta_{\mu \bar{\sigma} \bar{\theta}}^{\mu' \bar{\sigma}' \bar{\theta}'}=\delta_{\mu}^{\mu'} \delta_{\bar{\sigma} \bar{\theta}}^{\bar{\sigma}' \bar{\theta}'}$, where $\delta_{\bar{\sigma} \bar{\theta} }^{\bar{\sigma}' \bar{\theta}'}=\delta_{(\bar{\sigma} \bar{\theta}) (\bar{\sigma}' \bar{\theta}')}=\frac{1}{\hat{\bold{E}}}\delta(\bar{\sigma}-\bar{\sigma}') \delta^2(\bar{\theta}-\bar{\theta}')$.

%--------section3-------------%
\section{Consistent truncations to type IIA and IIB supergravities}
In this section, we will show that we can consistently truncate the string geometry model eq.~(\ref{action of bos string-geometric model}) to the type IIA and IIB supergravities if we apply appropriate configurations to the model, respectively  and take $\alpha' \to 0$.

From eq.~(\ref{action of bos string-geometric model}), the equations of motion of $\mathbf{G}_{\bold{I}\bold{J}}$, $\Phi$, $\mathbf{B}_{\bold{L}_1\bold{L}_2}$ and $\bold{A}_{\bold{L}_1 \cdots \bold{L}_{p-1}}$ are derived as
\begin{eqnarray}
   && \bold{R}_{\bold{I}\bold{J}} + 2 \nabla_{\bold{I}} \nabla_{\bold{J}} \Phi -\frac{1}{4} \bold{H}_{\bold{I}\bold{L}_{1}\bold{L}_{2}} \bold{H}_{\bold{J}}^{\bold{L}_{1}\bold{L}_{2}}-\frac{1}{2} \bold{G}_{\bold{I}\bold{J}} (\bold{R}+4 \nabla_{\bold{I}} \nabla^{\bold{I}} \Phi -4 \partial_{\bold{I}} \Phi \partial^{\bold{I}} \Phi -\frac{1}{2} |\bold{H}_{3}|^{2})
\nonumber \\
&& -\frac{1}{2} e^{2 \Phi} \sum_{p=1}^9 \Big[ \frac{1}{(p-1)!} \tilde{\bold{F}}_{\bold{I} \bold{L}_{1} \cdots \bold{L}_{p-1}} \tilde{\bold{F}}_{\bold{J}}^{\bold{L}_{1} \cdots \bold{L}_{p-1}} -\frac{1}{2} \bold{G}_{\bold{I}\bold{J}} |\tilde{\bold{F}}_{p}|^{2}  \Big]=0,  \label{G}\\
    && \bold{R}+4 \nabla_{\bold{I}} \nabla^{\bold{I}} \Phi -4 \partial_{\bold{I}} \Phi \partial^{\bold{I}} \Phi -\frac{1}{2} |\bold{H}_{3}|^{2}=0,   \label{Phi}\\ 
&&
\sum_{p=3}^9 \sum_{n=0}^{[\frac{p-3}{2}]} \frac{1}{2^{n+1} \cdot (p-2)!}
 \bold{F}_{ \bold{I}_{1} \cdots \bold{I}_{p-2-2n} } 
\bold{B}_{\bold{J}_{1} \bold{K}_{1} } \cdots \bold{B}_{\bold{J}_{n} \bold{K}_{n} } 
 \tilde{\bold{F}}^{ \bold{I}_{1} \cdots \bold{I}_{p-2-2n} \bold{J}_{1} \bold{K}_{1}  \cdots \bold{J}_{n} \bold{K}_{n}  \bold{L}_{1}\bold{L}_{2} } \nonumber \\
 &&+\nabla_{\bold{I}} ( e^{-2\Phi} \bold{H}^{\bold{I} \bold{L}_{1} \bold{L}_{2} }) =0,  \label{B} \\
&& \nabla_{\bold{I}} \tilde{\bold{F}}^{\bold{I} \bold{L}_{1} \cdots \bold{L}_{p-1}} + \left( \frac{1}{2} \right)^{n} \nabla_{\bold{I}} \Big[  \bold{B}_{\bold{J}_{1} \bold{K}_{1} } \cdots \bold{B}_{\bold{J}_{n} \bold{K}_{n} } \tilde{\bold{F}}^{\bold{J}_{1}\bold{K}_{1} \cdots \bold{J}_{n}\bold{K}_{n} \bold{I} \bold{L}_{1} \cdots \bold{L}_{p-1} }   \Big]=0, \label{A}
\end{eqnarray}
respectively.

We consider particular configurations, which we call IIA and IIB string background configurations, 
\begin{description}
\item{Metric:}
\begin{eqnarray}
&&\mathbf{G}_{00} \left(\bar{\tau}, \bold{X} \right) = -1 \nonumber \\ 
&&\mathbf{G}_{(\mu_{1} \bar{\sigma}_{1} \bar{\theta}_1)(\mu_{2} \bar{\sigma}_{2} \bar{\theta}_2)}\left(\bar{\tau}, \bold{X} \right)
=
G_{\mu_{1} \mu_{2}} \left( \bold{X}(\bar{\sigma}_{1} \bar{\theta}_1) \right) \delta_{(\bar{\sigma}_{1} \bar{\theta}_1) (\bar{\sigma}_{2} \bar{\theta}_2)  }\delta_{(\bar{\sigma}_{1} \bar{\theta}_1) (\bar{\sigma}_{1} \bar{\theta}_1) }  
\nonumber \\
&&\text{the others} = 0,  \label{Gansatz}
\end{eqnarray}
\item{Scalar field:}
\begin{eqnarray}
\Phi\left(\bar{\tau},\bold{X}\right)=\int d \bar{\sigma} 
d^2 \bar{\theta} \hat{\bold{E}} \delta_{(\bar{\sigma}, \bar{\theta}) (\bar{\sigma},\bar{\theta})}
\, \phi(\bold{X} (\bar{\sigma}, \bar{\theta}))
\label{Sansatz}
\end{eqnarray}
\item{B field:}
\begin{eqnarray}
&&\mathbf{B}_{(\mu_{1} \bar{\sigma}_{1} \bar{\theta}_1)(\mu_{2} \bar{\sigma}_{2} \bar{\theta}_2)}\left(\bar{\tau}, \bold{X} \right)
=
B_{\mu_{1} \mu_{2}} \left( \bold{X}(\bar{\sigma}_{1} \bar{\theta}_1) \right) \delta_{(\bar{\sigma}_{1} \bar{\theta}_1) (\bar{\sigma}_{2} \bar{\theta}_2)}\delta_{(\bar{\sigma}_{1} \bar{\theta}_1) (\bar{\sigma}_{1} \bar{\theta}_1) }\nonumber \\
&&\text{the others} = 0,  \label{Bansatz}
\end{eqnarray}
\item{p-form field:}
\begin{eqnarray}
&&\mathbf{A}_{(\mu_{1} \bar{\sigma}_{1} \bar{\theta}_1 )\cdots(\mu_{p} \bar{\sigma}_{p} \bar{\theta}_p)}\left(\bar{\tau}, \bold{X} \right) = A_{\mu_{1} \cdots \mu_{p}} \left( \bold{X}(\bar{\sigma}_{1} \bar{\theta}_1) \right) \delta_{(\bar{\sigma}_{1} \bar{\theta}_1) (\bar{\sigma}_{2} \bar{\theta}_2)}
\cdots
\delta_{(\bar{\sigma}_{p-1} \bar{\theta}_{p-1}) (\bar{\sigma}_{p} \bar{\theta}_p)}
\delta_{(\bar{\sigma}_{1} \bar{\theta}_1) (\bar{\sigma}_{1} \bar{\theta}_1)}, 
\nonumber \\
&&\text{the others} = 0, \label{pansatz}
\end{eqnarray}
\end{description}
where
\begin{eqnarray}
&&A_{\mu_{1} \cdots \mu_{p}}=0 \mbox{  (p: even)}
\nonumber \\
&&\tilde{F}_8=-* \tilde{F}_2,
\nonumber \\
&&\tilde{F}_6=* \tilde{F}_4,
\nonumber \\
&&A_1=C_1,
\nonumber \\
&&A_3=C_3+B_2 \wedge C_1,
\label{IIAconfiguration}
\end{eqnarray}
for IIA string background configuration, or
\begin{eqnarray}
&&A_{\mu_{1} \cdots \mu_{p}}=0 \mbox{  (p: odd)}
\nonumber \\
&&\tilde{F}_9=* \tilde{F}_1,
\nonumber \\
&&\tilde{F}_7=-* \tilde{F}_3,
\nonumber \\
&&\tilde{F}_5=* \tilde{F}_5,
\nonumber \\
&&A_0=C_0,
\nonumber \\
&&A_2=C_2+B_2 C_0,
\nonumber \\
&&A_4=C_4+\frac{1}{2}B_2 \wedge C_2 + \frac{1}{2} B_2 \wedge B_2 C_0,
\label{IIBconfiguration}
\end{eqnarray}
for IIB string background configuration.
$G_{\mu_{1} \mu_{2}} \left( x \right)$ is a symmetric tensor field, $\phi \left( x \right)$ is a scalar field, $B_{\mu_{1} \mu_{2}} \left( x \right)$ is an B field and  $C_{\mu_{1} \cdots \mu_{p}}\left( x \right)$ are p-form fields on a 10-dimensional spacetime. We  will show that the IIA and IIB configurations satisfy the equations of motion of the string geometry model in $\alpha' \to 0$ if and only if the 10-dimensional fields satisfy the equations of motion of the IIA and IIB supergravities, respectively.

We remark that the string background configuration has a non-trivial dependence on the worldsheet. The consistent truncation will be ensured due to the relation between the worldsheet dependence of the fields and of the indices of the string geometry fields. For example, see $(\bar{\sigma}_1 \bar{\theta}_1)$ dependence on the string background configuration for the metric. In addition, the factor $\delta_{(\bar{\sigma}_{1} \bar{\theta}_1) (\bar{\sigma}_{1} \bar{\theta}_1)}$ reflects that the point particle limit is a field theory. 

The $\alpha' \to 0$ limits of the equations of motions ~(\ref{G})~$\sim$~(\ref{A}) with the IIA string background configuration are equivalent to the equations of motion of the type IIA supergravity
\begin{eqnarray}
S_{IIA}& =& \frac{1}{2 \kappa^{2}_{10}} \Biggl( \int d^{10}x \sqrt{-G}  \left( e^{ -2 \phi } \left( R + 4 \nabla_{\mu} \phi \nabla^{\mu} \phi - \frac{1}{2} |H|^{2}  \right)
- \frac{1}{2} |\tilde{F}_2|^{2}- \frac{1}{2} |\tilde{F}_4|^{2} \right) \nonumber \\
&&-\frac{1}{2}\int B \wedge dC_3 \wedge dC_3
\Biggr).
\label{IIAaction}
\end{eqnarray}
The $\alpha' \to 0$ limits of the equations of motions ~(\ref{G})~$\sim$~(\ref{A}) with the IIB string background configuration are also equivalent to the equations of motion of the type IIB supergravity,
\begin{eqnarray}
S_{IIB}& =& \frac{1}{2 \kappa^{2}_{10}} \Biggl( \int d^{10}x \sqrt{-G}  \left( e^{ -2 \phi } \left( R + 4 \nabla_{\mu} \phi \nabla^{\mu} \phi - \frac{1}{2} |H|^{2}  \right)
- \frac{1}{2} |\tilde{F}_1|^{2}- \frac{1}{2} |\tilde{F}_3|^{2} - \frac{1}{4} |\tilde{F}_5|^{2} \right) \nonumber \\
&&-\frac{1}{2}\int C_4 \wedge H \wedge dC_2
\Biggr),
\label{IIBaction}
\end{eqnarray}
and the self-dual condition,
\begin{eqnarray}
\tilde{F}_5=* \tilde{F}_5.
\nonumber
\end{eqnarray}

Here we display a mechanism how the $\alpha' \to 0$ limit of the equation of motion of $\bold{G}_{(\mu \bar{\sigma}_{1} \bar{\theta}_1) (\nu \bar{\sigma}_{2} \bar{\theta}_2)}$ with the string background configuration is equivalent to the equation of motion of $G_{\mu \nu}$.
By substituting the string background configuration,
the left hand side of the Einstein equation becomes
\begin{eqnarray}
&&\bold{R}_{(\mu \bar{\sigma}_1 \bar{\theta}_1) (\nu \bar{\sigma}_2 \bar{\theta}_2)} 
-\frac{1}{2}\bold{G}_{(\mu \bar{\sigma}_1 \bar{\theta}_1) (\nu \bar{\sigma}_2 \bar{\theta}_2)} \bold{R}
\nonumber \\
&=&
\delta_{(\bar{\sigma}_1 \bar{\theta}_1) (\bar{\sigma}_2 \bar{\theta}_2)}
\delta_{(\bar{\sigma}_1 \bar{\theta}_1)  (\bar{\sigma}_1 \bar{\theta}_1)}
\biggl(R_{\mu \nu} \left( \bold{X}_{\hat{D}_T}(\bar{\sigma}_{1} \bar{\theta}_1) \right)
\nonumber \\
&&-\frac{1}{2}
G_{\mu \nu} \left( \bold{X}_{\hat{D}_T}(\bar{\sigma}_{1} \bar{\theta}_1)  \right)
\int d \bar{\sigma} d^2 \bar{\theta} \hat{\bold{E}}(\bar{\sigma} \bar{\theta})   \delta_{(\bar{\sigma} \bar{\theta}) (\bar{\sigma} \bar{\theta})} 
R \left( \bold{X}_{\hat{D}_T}(\bar{\sigma} \bar{\theta}) \right)  \biggr).
\end{eqnarray}
As one can see in this formula, if an equation of motion includes a trace (in $\bold{R}$ in this case), the reduced equation of motion includes an extra summation  $\int d \bar{\sigma} d^2 \bar{\theta} \hat{\bold{E}}(\bar{\sigma} \bar{\theta})   \delta_{(\bar{\sigma} \bar{\theta}) (\bar{\sigma} \bar{\theta})}$ against the equation of motion of the string backgrounds. Fortunately,  the terms including the extra summation vanish by using the string background configuration and the equation of motion of the scalar, as one can see below. Actually, by substituting the string background configuration into the equation of motion of $\bold{G}_{(\mu \bar{\sigma}_{1} \bar{\theta}_1) (\nu \bar{\sigma}_{2} \bar{\theta}_2)}$, we obtain
\begin{eqnarray}
0&=&  \bold{R}_{(\mu \bar{\sigma}_{1} \bar{\theta}_1) (\nu \bar{\sigma}_{2} \bar{\theta}_2)} - \frac{1}{4} \bold{H}_{(\mu \bar{\sigma}_{1} \bar{\theta}_1) \bold{I}_{1} \bold{I}_{2}}  \bold{H}_{(\nu \bar{\sigma}_{2} \bar{\theta}_2)}^{\bold{I}_{1} \bold{I}_{2}} + 2 \nabla_{(\mu \bar{\sigma}_{1} \bar{\theta}_1)} \nabla_{(\nu \bar{\sigma}_{2} \bar{\theta}_2)} \Phi 
\nonumber \\
&&
-\frac{1}{2} \bold{G}_{(\mu \bar{\sigma}_{1} \bar{\theta}_1) (\nu \bar{\sigma}_{2} \bar{\theta}_2)}  (\bold{R}+4 \nabla_{\bold{I}} \nabla^{\bold{I}} \Phi -4 \partial_{\bold{I}} \Phi \partial^{\bold{I}} \Phi -\frac{1}{2} |\bold{H}_{3}|^{2})
\nonumber \\
&&-\frac{1}{2}e^{2\Phi}
\sum_{ p=1}^{9 } \Big( \frac{1}{ (p-1)! }  
\tilde{\bold{F}}_{(\mu \bar{\sigma}_{1} \bar{\theta}_1)\bold{I}_{1} \cdots \bold{I}_{p-1}}   \tilde{\bold{F}}_{ (\nu \bar{\sigma}_{2} \bar{\theta}_2)}^{ \bold{I}_{1} \cdots \bold{I}_{p-1}}  -\frac{1}{2} \bold{G}_{(\mu \bar{\sigma}_{1} \bar{\theta}_1) (\nu \bar{\sigma}_{2} \bar{\theta}_2)}
|\tilde{\bold{F}}_{p}|^{2} \Big) 
\nonumber \\
&=& \delta_{(\bar{\sigma}_{1} \bar{\theta}_1) (\bar{\sigma}_{2} \bar{\theta}_2)  }\delta_{(\bar{\sigma}_{1} \bar{\theta}_1) (\bar{\sigma}_{1} \bar{\theta}_1) } 
\Biggl( R_{\mu \nu}\left(  \bold{X} (\bar{\sigma}, \bar{\theta}) \right) - \frac{1}{4} H_{\mu \mu_{1} \mu_{2}}  H_{\nu}^{\mu_{1} \mu_{2}} + 2 \nabla_{\mu } \nabla_{ \nu } \phi 
\nonumber \\
&&-\frac{1}{2}e^{2\int  d \bar{\sigma} 
d^2 \bar{\theta} \hat{\bold{E}} \delta_{(\bar{\sigma}, \bar{\theta})  (\bar{\sigma}, \bar{\theta}) } 
 \phi(\bold{X} (\bar{\sigma}, \bar{\theta}))}
\sum_{ p=1}^{9 } \frac{1}{ (p-1)! }  \tilde{F}_{\mu \mu_{1} \cdots mu_{p-1}}   \tilde{F}_{\nu}^{ \mu_{1} \cdots \mu_{p-1}}  
\nonumber \\
&&-\frac{1}{2} G_{\mu \nu} 
\int  d \bar{\sigma} 
d^2 \bar{\theta} \hat{\bold{E}} \delta_{(\bar{\sigma}, \bar{\theta})  (\bar{\sigma}, \bar{\theta}) } 
\sum_{ p=1}^{9} 
|\tilde{F}_{p} \left(\bold{X} (\bar{\sigma}, \bar{\theta}) \right)  |^{2} \Biggr).
\label{reduction}
\end{eqnarray}
In the second equality, we have used the equation of motion of the scalar (\ref{Phi}) and the terms in the second line, which includes a part of the extra summation,  vanishes. By using the common property in the IIA and IIB string background configurations, 
\begin{eqnarray}
\tilde{F}_{10-n}= \pm *\tilde{F}_{n}
\nonumber
\end{eqnarray}
we obtain 
\begin{eqnarray}
|\tilde{F}_{10-n}|^{2}=-|\tilde{F}_{n}|^{2}. 
\label{AbsoluteRelation}
\end{eqnarray}
If we substitute this relation into (\ref{reduction}), the last term, which includes the remaining part of the extra summation, vanishes. Furthermore, If we take  
\begin{eqnarray}
\alpha' \to 0,
\nonumber
\end{eqnarray}
we obtain
\begin{equation}
e^{2\int  d \bar{\sigma} 
d^2 \bar{\theta} \hat{\bold{E}} \delta_{(\bar{\sigma}, \bar{\theta})  (\bar{\sigma}, \bar{\theta}) } 
 \phi(\bold{X} (\bar{\sigma}, \bar{\theta}))}
\to
e^{2\phi(x)}, \label{limit}
\end{equation}
and thus,  (\ref{reduction}) gets to be equivalent to
\begin{eqnarray}
0=R_{\mu \nu}\left(x \right) - \frac{1}{4} H_{\mu \mu_{1} \mu_{2}}  H_{\nu}^{\mu_{1} \mu_{2}} + 2 \nabla_{\mu } \nabla_{ \nu } \phi 
-\frac{1}{2}e^{2\phi} 
\sum_{ p=1}^{9 } \frac{1}{ (p-1)! }  \tilde{F}_{\mu \mu_{1} \cdots mu_{p-1}}   \tilde{F}_{\nu}^{ \mu_{1} \cdots \mu_{p-1}} .
\label{reducedEOM}
\end{eqnarray}
This formula is equivalent to the equation of motion of the metric of 
\begin{eqnarray}
S' = \frac{1}{2 \kappa^{2}_{10}} \left( \int d^{10}x \sqrt{-G}  \left( e^{ -2 \phi } \left( R + 4 \nabla_{\mu} \phi \nabla^{\mu} \phi - \frac{1}{2} |H|^{2}  \right)
- \frac{1}{2} \sum_{p=1}^9|\tilde{F}_{p}|^{2}\right)
\right)
\label{NewAction}
\end{eqnarray}
under the equation of motion of the scalar of this action and  (\ref{AbsoluteRelation}).  The same applies to the other fields.
 Furthermore,  the equations of motion of  (\ref{NewAction}) where the zero mode part of the IIA or IIB string background configurations (\ref{IIAconfiguration}), (\ref{IIBconfiguration}) are substituted are equivalent to the equations of motion of IIA or IIB supergravities (\ref{IIAaction}),  (\ref{IIBaction}), as one can see a proof in an appendix of \cite{Fukuma:1999jt}.   
 Thus, the IIA and IIB supergravities, which possess the Chern-Simons terms, are derived from the string geometry model,  which does not possess the Chern-Simons term.  
 Therefore, we conclude that the string backgrounds can be embedded into the string geometry model in the sense of the consistent truncation in $\alpha' \to 0$. 
 
 In the NS-NS (bosonic) sector of the string geometry theory,  the above discussion is valid without taking $\alpha' \to 0$ limit as in \cite{Honda:2020sbl}.  We can see this from the fact that the last two lines are absent in (\ref{reduction}) in the NS-NS sector, for example. This fact will be important to derive the non-linear sigma model on the NS-NS backgrounds as we mentioned in the introduction\footnote{We do not need the special mechanism that the extra summation vanishes as in (\ref{reduction}) to derive the supergravities because the extra summation automatically vanishes in $\alpha' \to 0$ as in  (\ref{limit}).  Howerver,  the special mechanism is necessary to derive a non-linear sigma model around  NS-NS background.}.  

We could not reproduce  the R-R sector of the supergravities from string geometry theory without $\alpha' \to 0$ limit. This suggests that  string geometry theory cannot reproduce non-linear sigma models on R-R backgrounds in the NS-R formalism. This is consistent with the fact that it is too difficult to formulate  non-linear sigma models on R-R backgrounds in the NS-R formalism.

%--------section4-------------%
\section{Equations that determine a string background}
Because the string background configuration ~(\ref{Gansatz})~$\sim$~(\ref{IIBconfiguration}) is stationary with respect to the string geometry time $\bar{\tau}$, the energy of it is defined as
\begin{eqnarray}
E&=&\int \mathcal{D} E  \mathcal{D} \bold{X} T_{00}
\nonumber \\
&=&\int \mathcal{D} E  \mathcal{D} \bold{X}
(
-2 \nabla_{0} \nabla_{0} \Phi 
+\frac{1}{4} \mathbf{H}_{0L_{1} L_{2}} \mathbf{H}_{0}^{L_{1}L_{2}}
+\frac{1}{2}e^{2\Phi}
\sum_{ p=1}^{9 } \frac{1}{ (p-1)! }  
\tilde{\bold{F}}_{0 \bold{I}_{1} \cdots \bold{I}_{p-1}}   \tilde{\bold{F}}_{ 0}^{ \bold{I}_{1} \cdots \bold{I}_{p-1}} \nonumber \\
&&+G_{00}(-2 \nabla_{I} \Phi \nabla^{I} \Phi + 2 \nabla_{I} \nabla^{I} \Phi - \frac{1}{4} | \mathbf{H} |^{2}-\frac{1}{4}e^{2 \Phi}\sum_{p=1}^9
|\tilde{\bold{F}}_p|^2)     
) \nonumber \\
&=&\int \mathcal{D} \bold{X} \int \mathcal{D} E \int  d \bar{\sigma} 
d^2 \bar{\theta} \hat{\bold{E}} \delta_{(\bar{\sigma}, \bar{\theta})  (\bar{\sigma}, \bar{\theta}) } 
 (2 \nabla_{\mu} \phi \left( \bold{X} (\bar{\sigma}, \bar{\theta}) \right) \nabla^{\mu} \phi  - 2 \nabla_{\mu} \nabla^{\mu} \phi + \frac{1}{4} | H |^{2})  
\nonumber \\
&=&\int \mathcal{D} X \int \mathcal{D} e \int d \bar{\sigma} \bar{e}(\bar{\sigma})   \delta_{\bar{\sigma} \bar{\sigma}} 
(2 \nabla_{\mu} \phi \left( X(\bar{\sigma}) \right) \nabla^{\mu} \phi  - 2 \nabla_{\mu} \nabla^{\mu} \phi + \frac{1}{4} | H |^{2})  
\nonumber \\
&=&\int d^{10}x \sqrt{-G(x)}
(2 \nabla_{\mu} \phi(x)\nabla^{\mu} \phi  - 2 \nabla_{\mu} \nabla^{\mu} \phi +\frac{1}{4} | H |^{2})
\label{energy}
\end{eqnarray}
On the second and third line in the above formula, we have substituted ~(\ref{Gansatz})~$\sim$~(\ref{IIBconfiguration}) and obtained the fourth line. On the fourth line, because $ \delta_{(\bar{\sigma}, \bar{\theta})  (\bar{\sigma}, \bar{\theta}) }  \propto \bar{\theta}\bar{\bar{\theta}}$, if one integrates  $\bar{\theta}$ and $\bar{\bar{\theta}}$, only the bosonic leading terms remain and we obtained the fifth line.  On the fifth line, we have regularized the integral over the embedding function as
\begin{eqnarray}
&&\int \mathcal{D} X=\prod_{j=1}^N \int d^{10}x_j \sqrt{-G(x_j)}
\nonumber \\
&&\int d^{10}x_j \sqrt{-G(x_j)}=1
\nonumber \\
&&\int \mathcal{D} e \int d \bar{\sigma} \bar{e}(\bar{\sigma})   \delta_{\bar{\sigma} \bar{\sigma}} = \frac{1}{N}\sum_{i=1}^N,
\end{eqnarray}
and obtained the sixth line.

Therefore, in string geometry theory, a string background is determined by minimizing the energy (\ref{energy}) of the solutions to the IIA or IIB equations of motions. In other words, in the IIA case, by using the method of Lagrange multiplier, the equations that determine string backgrounds are obtained by differentiating 
\begin{eqnarray}
\tilde{E}
&=&
E
+\int d^{10}x \sqrt{-G(x)}
(\lambda_G^{\mu \nu}(x) f^G_{\mu \nu}(x)
+\lambda_{\phi}(x) f^{\phi}(x)
+\lambda_B^{\mu \nu}(x) f^B_{\mu \nu}(x)
\nonumber \\
&&
\qquad \qquad \qquad \qquad \quad +\lambda_{C_1}^{\mu}(x) f^{C_1}_{\mu}(x)
+\lambda_{C_{3}}^{\mu_1 \mu_2 \mu_{3}}(x) f^{C_{3}}_{\mu_1 \mu_2 \mu_{3}}(x))
\end{eqnarray}
with respect to 
the IIA string backgrounds $G_{\mu \nu}(x)$, $\phi(x), B_{\mu \nu}(x)$, $C_1(x)$ and $C_3(x)$
and
the Lagrange multipliers $\lambda_G^{\mu \nu}(x)$, 
$\lambda_{\phi}(x)$,
$\lambda_B^{\mu \nu}(x)$, 
$\lambda_{C_1}^{\mu}(x)$ and
$\lambda_{C_{3}}^{\mu_1 \mu_2 \mu_{3}}(x)$, 
where 
$ f^G_{\mu \nu}(x)=0$, $f^{\phi}(x)=0$, $f^B_{\mu \nu}(x)=0$,  $f^{C_1}_{\mu}(x)=0$ and $f^{C_{3}}_{\mu_1 \mu_2 \mu_{3}}(x)=0$
represent the IIA equations of motions, respectively. The same applies to the IIB case.

%--------section5-------------%
\section{Consistent truncations to heterotic and type I supergravities}
Let us generalize the model (\ref{action of bos string-geometric model}) and consider
\begin{eqnarray}
S=\int \mathcal{D}\bold{E} \mathcal{D}\bar{\tau} \mathcal{D}\bold{X}_{\hat{D}} 
\sqrt{\bold{G}} \left( e^{-2 \Phi} \left( \mathbf{R}  + 4 \nabla_{\bold{I}} \Phi \nabla^{\bold{I}} \Phi - \frac{1}{2} |\tilde{\mathbf{H}} |^{2} -\mbox{tr}(|\mathbf{F}|^2)  \right) -\frac{1}{2}\sum_{p=1}^9
|\tilde{\bold{F}}_p|^2     
\right),
\label{extended action}
\end{eqnarray}
where 
$\tilde{\mathbf{H}}=d\mathbf{B}-\boldsymbol{\omega}_3$, 
$\boldsymbol{\omega}_3=\mbox{tr}(\mathbf{A}\wedge d\mathbf{A} -\frac{2i}{3}\mathbf{A} \wedge \mathbf{A} \wedge \mathbf{A})$, and 
$\mathbf{A}$ is a $N \times N$ Hermitian gauge field, whose field strength is given by $\mathbf{F}$. 
$\mathbf{A}$ can be consistently truncated to 0 in this model and the model  (\ref{action of bos string-geometric model})  is obtained. Thus, the IIA and IIB supergravities are derived by consistent truncations of the model (\ref{extended action}).

On the other hand,  the heterotic supergravity 
\begin{eqnarray}
S_{het}& =& \frac{1}{\kappa^{2}_{10}}\int d^{10}x \sqrt{-G}  e^{ -2 \phi } \left( R + 4 \nabla_{\mu} \phi \nabla^{\mu} \phi - \frac{1}{2} |\tilde{H}|^{2}   - \frac{\kappa^{2}_{10}}{g_{10}^2} \mbox{tr}(|F|^2) \right),
\label{hetero action}
\end{eqnarray}
where $\tilde{H}=dB-  \frac{\kappa^{2}_{10}}{g_{10}^2}\omega_3$,
is derived by consistently truncating the model  (\ref{extended action}) as in the section 3.  Here, we have introduced the constants $\kappa_{10}$ and  $g_{10}$ by shifting the dilaton and rescaling in the 10 dimensions. The gauge field can be truncated to the adjoint representation of $SO(32)$ and $E_8 \times E_8$. 

The heterotic supergravity can be shown to be equivalent to the type I supergravity
\begin{eqnarray}
S_{I}& =& \frac{1}{\kappa^{2}_{10}}\int d^{10}x \sqrt{-G}  \left( e^{ -2 \phi }  \left(R + 4 \nabla_{\mu} \phi \nabla^{\mu} \phi \right)- \frac{1}{2} |\tilde{F}_3|^{2}\right)   -\frac{1}{g^{2}_{10}}\int d^{10}x \sqrt{-G}  e^{ - \phi } \mbox{tr}(|F|^2), 
\nonumber \\
\label{type I action}
\end{eqnarray}
where $\tilde{F}_3=dC_2-  \frac{\kappa^{2}_{10}}{g_{10}^2}\omega_3$, by a transformation: $B \to C_2$, $G_{\mu \nu} \to e^ {- \phi}G_{\mu \nu}$ and  $\phi \to -\phi$. The gauge field can be truncated to the adjoint representation of $SO(32)$. As a result,  the supergravities on all the ten-dimensional vacua in string theory, are derived from the single model  (\ref{extended action}).

\section{Conclusion and Discussion}
In this paper, we have shown that the supergravities on all the ten-dimensional vacua in superstring theory, are derived from a single string geometry model by consistent truncations. That is, arbitrary configurations of all the type IIA, IIB, SO(32) type I, and   SO(32)  and $E_8 \times E_8$ heterotic superstring backgrounds are embedded in configurations of fields of a single string geometry model.
Especially,  the single action of the string geometry model is consistently truncated to the supergravity actions by  applying the  corresponding superstring background configurations to the model, respectively  and take $\alpha' \to 0$. That is, the $\alpha' \to 0$ limits of the equations of motions with those string background configurations are equivalent to the equations of motion of the corresponding supergravities, respectively. This means that classical dynamics of all the type IIA, IIB, SO(32) type I, and   SO(32)  and $E_8 \times E_8$ heterotic  superstring backgrounds are described as a part of classical dynamics in string geometry theory. This fact supports the conjecture that string geometry theory is a non-perturbative formulation of string theory.

The above results are consistent with the fact that one can derive both the type IIA and IIB perturbative string theories on the flat background from a single string geometry model as shown in  \cite{Sato:2017qhj}:  in this case, the configurations of the backgrounds are formally the same, whereas the charts that cover the backgrounds are different (IIA and IIB charts).  These results strongly indicate that string geometry theory does not depend on string backgrounds. Here we comment on supersymmetry.  Although  the single action of the ten-dimensional gravity (\ref{NewAction}) possesses both fields in type IIA and IIB supergravities, namely all the R-R filelds with odd and even degrees,  the action cannot be generalized to be supersymmetric even if fermions are coupled. However, in string geometry theory, an arbitrary action can be generalized to
%possess a supersymmetry, whose origin corresponds to the world-sheet supersymmetry (\ref{supertrans})
be supersymmetric, as one can see in \cite{Sato:2017qhj}. Actually, the string geometry models (\ref{action of bos string-geometric model}) and (\ref{extended action}) are supersymmetric, although they possess the tensor fields that include all the R-R fields with odd and even degrees. 

Furthermore, we have defined an energy of the superstring background configurations, because they are stationary with respect to the string geometry time $\bar{\tau}$. Thus, a superstring background can be determined by minimizing the energy of the solutions to the equations of motions of the superstring backgrounds. Therefore, we conclude that string geometry theory includes a non-perturbative effect that determines a superstring background. 

%We will derive the path-integral of a non-linear sigma model on a NS-NS background from fluctuations around ~(\ref{Gansatz})~$\sim$~(\ref{IIBconfiguration}) when the R-R backgrounds are turned off. We already showed that this is true when the string background is flat \cite{Sato:2017qhj, Sato:2019cno}. 

%--------section Ack-------------% 
\section*{Acknowledgements}
We would like to thank  
K. Hashimoto,
Y. Hyakutake,
Y. Nakayama
T. Onogi,
S. Sugimoto,
Y. Sugimoto,
S. Yamaguchi,
and especially 
H. Kawai and A. Tsuchiya
for long and valuable discussions.

%------------BIBLIOGRAPHY--------------%

\bibliographystyle{prsty}

\end{document}